UDC 004.853:373

**Valentyna V. Kovalenko**
PhD of Pedagogical Sciences,
Senior Researcher of the Department of Cloud-Oriented Systems of Education Informatization
Institute of Information Technologies and Learning Tools of NAES of Ukraine, Kyiv, Ukraine
ORCID ID 0000-0002-4681-5606
*valyavako@gmail.com*

**Maiia V. Marienko**
PhD of Pedagogical Sciences,
Senior Researcher of the Department of Cloud-Oriented Systems of Education Informatization
Institute of Information Technologies and Learning Tools of NAES of Ukraine, Kyiv, Ukraine
ORCID ID0000-0002-8087-962X
*popelmayal@gmail.com*

**Alisa S. Sukhikh**
PhD of Pedagogical Sciences,
Senior Researcher of the Department of Cloud-Oriented Systems of Education Informatization
Institute of Information Technologies and Learning Tools of NAES of Ukraine, Kyiv, Ukraine
ORCID ID 0000-0001-8186-1715
*alisam@ukr.net*

# USE OF AUGMENTED AND VIRTUAL REALITY TOOLS IN A GENERAL SECONDARY EDUCATION INSTITUTION IN THE CONTEXT OF BLENDED LEARNING

**Abstract.** The study examines the problem of using augmented and virtual reality in the process of blended learning in general secondary education. Analysis of recent research and publications has shown that the use of augmented and virtual reality in the educational process has been considered by scientists. However, the target group in these studies is students of higher education institutions. Most of the works of scientists are devoted to the problem of introducing augmented reality into the traditional educational process. At the same time, the use of augmented and virtual reality technologies in the process of blended learning remains virtually unexplored. The study analyzes the meaning of the concept of "blended learning". The conceptual principles of blended learning are considered. It has been found that scholars differ in their understanding of the concept of "blended learning". Sometimes researchers distinguish between the components of blended learning: full-time and online learning. The study presents the special advantages of blended learning and the taxonomy of blended learning. It was found that there are some difficulties in implementing blended learning. The article outlines the practical use of virtual and augmented reality. The definition of augmented and virtual reality is given. The mixed reality is considered as a separate kind of notion. Separate applications of virtual and augmented reality that can be used in the process of blended learning are considered (MEL Chemistry VR; Anatomyou VR; Google Expeditions; EON-XR). As a result of the study, the authors propose possible ways to use augmented reality in the educational process. The model of using augmented and virtual reality in blended learning in general secondary education institutions was designed. It consists of the following blocks: goal; teacher's activity; forms of education; teaching methods; teaching aids; organizational forms of education; pupil activity and results. Based on the model, the methodology of using augmented and virtual reality in blended learning in general secondary education was developed. The methodology contains the following components: target component, content component, technological component and resultant component. The methodology is quite universal and can be used for any subject in general secondary education. The types of lessons in which it is expedient to use augmented (AR) and virtual reality(VR) are determined. Recommendations are given at which stage of the lesson it is better to use AR and VR tools (depending on the type of lesson).

## 1. INTRODUCTION

**Formulation of the problem.** Today, as the digital transformation of education has become a challenge to the educational system in Ukraine, the introduction of modern digital learning technologies and tools, in particular, the tools of augmented and virtual reality, is among the most crucial issues.The Ministry of Health of Ukraine has announced the criteria for the transition to distance learning during the Covid-19 pandemic. The "red" level of epidemic danger (high incidence) implies that general secondary schools stay open provided that at least 80% of their staff are vaccinated. Otherwise, education takes place in other forms, including distance learning.

"Hybrid learning", "distance learning", "distance education", "e-learning", "e-education", "blended learning", "distance learning" are the most common forms of learning during quarantine restrictions around the world. All of these terms are associated with the use of digital technology in teaching, but at the same time, "blended learning" is most commonly used. The need for the mass introduction of blended and distance learning led to their "forced" large-scale testing, which made it possible to identify their new features. Concepts and techniques of blended and distance learning, which were developed much earlier, are now useful. Cloud technologies, in particular means of communication, have made it possible to implement blended learning technologically.

For the current generation of general secondary school pupils, the educational process using the means of augmented and virtual reality is more understandable. Such tools are motivating pupils in mastering new knowledge, allow the teacher to organize the learning process, if necessary, adjust this process and monitor the success of each pupil [1]. Besides, the use of augmented and virtual reality tools provides an opportunity to track the nature of the pupil's mistakes and automatically determines the type of help needed. The teacher immediately notices the mistakes that the pupil makes. Thanks to the use of AR, VR and teamwork, errors can be seen immediately. Not only the teacher but also other pupils can help to correct the mistake. That is, the learning environment adapts to the level of pupils' knowledge, their needs, acquired knowledge, experience, and speed of perception of new material [2]. Teachers together with pupils will be able (according to the available tools) to determine jointly the sequence and pace of learning the material.

With the promotion of distance and blended learning, teachers are more interested [3] in new services for better perception of information by pupils. If cloud services are a tool for blended learning, then virtual and augmented reality will help to present information in an engaging way. Augmented reality is a tool for using and demonstrating theory during the lesson. The use of virtual and augmented reality applications will help the teacher to illustrate how theoretical knowledge can be used in practice. The teacher will be a tutor for pupils, providing direction to their learning trajectory. This will help pupils find the necessary knowledge on their own, be attentive to detail. The teacher and pupils act as members of the same team. The use of AR / VR is used to form students' practical skills in real conditions and life situations. It is easy to organize project activities and implement teamwork. Teachers can start with simpler topics and gradually focus on more complex ones. AR and VR provide an experience that pupils usually do not have access to. The latest technologies also play an important role in educating children with physical, social or cognitive impairments.

That is, we need to work on creating new teaching methods and improving the skills of older teachers. Limited resources in educational institutions still remain a serious problem. However, budgets for upgrading school equipment are gradually increasing. It is important that when choosing equipment, educational institutions pay attention to the scenarios of its use.

**Analysis of recent sources and publications.** Today, blended learning has become one of the most popular pedagogical concepts in education. It is a process that is supported by an effective combination of different teaching methods and learning styles. Although the concept itself appeared in the 1960s, it was not until the 2000s that it was studied in detail and actively implemented.

According to A. Stryuk, Y. Trius, V. Kukharenko [4], "blended learning" is a purposeful process of acquiring knowledge, skills and abilities in terms of integration of pupils' classroom and extracurricular educational activities based on implementation and mutual complementarity of traditional, electronic, distance and mobile learning technologies in the presence of pupils' self-control over the time, place, routes and pace of learning.

Researchers J. E. Prescott, K. Bundschuh, E. R. Kazakoff and P. Macaruso studied the implementation of a blended learning program for literacy in a kindergarten and up to the 5th grade in elementary school, including pupils learning English [5].

According to R. P. Murtikusuma, Hobri, A. Fatahillah, S. Hussen, R. R. Prasetyo, and M. A. Alfarisi, the organization of blended learning using Google Classroom will diversify the learning process, make it more engaging and enjoyable for both pupils and teachers [6].

The benefits described by L. B. Ni [7], which become available through computers and smartphones, are relevant to both school teachers and pupils. Blended classrooms, which use both traditional and technological teaching methods, have become the norm for many educators. Using Google Classroom gives pupils access to online learning.

I. Melnyk, N. Zaderei, and G. Nefodova[1] conducted a study of the main features and differences of virtual reality (VR), augmented reality (AR), and mixed reality (MR). The scientists have considered applications of augmented and virtual reality in modern educational process.

O. R. Oleksyuk considered the application of augmented reality technology in education. The study [2] revealed the meaning of the concept of augmented reality and considered the benefits of using augmented reality in the educational process. O. R. Oleksyuk analyzes the use of individual applications for different age groups of pupils and gives examples of the use of applications in teaching specific disciplines.

O. V. Syrovatskyi, S. O. Semerikov, Ye. O. Modlo, Yu. V. Yechkalo, and S. O. Zelinska performed a historical and technological analysis of the experience of using augmented reality tools for the development of interactive learning materials. In their study [8] they described the software for designing augmented reality tools for educational purposes.

V. V. Tkachuk, Yu. V. Echkalo, A. S. Taraduda, and I. P. Steblivets in the study [9] theoretically substantiated the feasibility of using augmented reality as a means of distance learning in quarantine. The scientists have identified a tool for visualizing laboratory equipment, namely the mobile application Electricity AR. In [10] the elements of the methodology of using the mobile application Electricity AR in the process of laboratory work were developed. Since augmented reality in this paper is considered in the process of distance learning, further research should focus on the feasibility of using augmented reality in the process of blended learning.

According to J. J. Stephan, A. S. Ahmed, A. H. Omran [11] there is a need to improve and develop the theory of blended learning through the use of a virtual reality environment to make it more effective. Pupils need to be able to experience the atmosphere of the lecture and overcome the difficulties that arise as a result of using blended or traditional learning. Research [11] found that pupils are most bored after 10 minutes of the lecture. Adding elements of virtual reality can add variety and thus make the lecture more interesting. In case of missing a lecture or the need to listen to it again, pupils can access the proposed website and view the lecture again using the virtual reality function.

In a study by G. Thorsteinsson and T. Page [12], blended learning is seen as a pedagogical approach. European educators in the FISTE Comenius 2.1 project have used a number of training measures to help improve in-service teacher education. The combination of lectures, visual diagrams, assessments and group classes today is the basis of learning in the classroom. Blended learning is a combination of all these approaches and the use of VR and AR.

In a study [13] K. Mumtaz, M. M. Iqbal, S. Khalid, T. Rafiq, S. M. Owais, M. Al Achhab compared the level of students' understanding in two scenarios: class lectures and lectures based on AR. The result showed that there is a difference between classroom teaching and AR teaching. The AR experience has a positive effect on student learning. In addition, students' confidence and motivation to learn are achieved.

The analysis of the existing research on the subject has shown that the idea of using the means of augmented and virtual reality in the process of blended learning in general secondary education remains practically unexplored. Most research relates to higher education and vocational education.

**The aim of the study.** The article aims to develop and substantiate the model of using the augmented and virtual reality tools in blended learning in general secondary education.

## 2. PRESENTATION OF THE MAIN MATERIAL

### 2.1. Conceptual principles of blended learning

Blended learning has become relevant and is actively developing, both in Ukraine and around the world, technologies of this teaching model are being introduced, so researchers are trying to reach a consensus in determining the characteristics of blended learning. Researchers define "blended learning" in different ways. Attempts by scientists to outline this concept demonstrate their different understanding of its content. Let us consider some of the existing views.

O. Krivonos and O. Korotun [14] understand blended learning as an educational concept in which the pupil acquires knowledge both independently (online) and in-person (with a teacher), which allows controlling the time, place, pace, and way of studying the material.

M. Oliver and K. Trigwell [15] note that blended learning has certain components, namely: a combination of face-to-face and online learning, technology, and methodology.

In their definitions, researchers believe that blended learning usually consists of two main components: face-to-face and online learning in different proportions. If we present the range of learning between offline and online, blended learning will be in the middle between them (see Fig. 1).

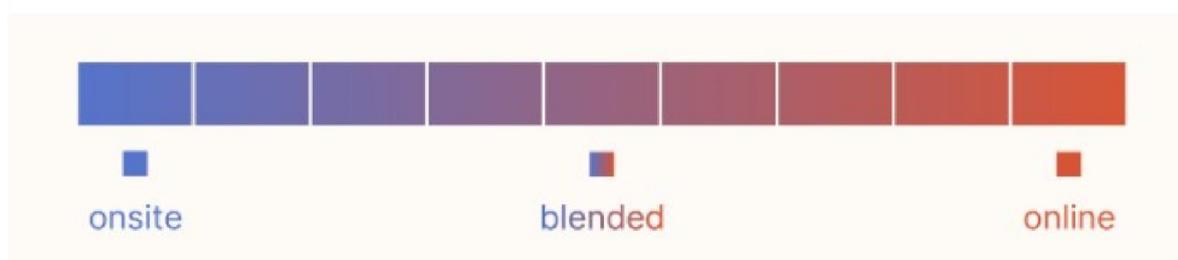

*Fig. 1. The range of learning between offline and online (for blended learning)[16]*

Blended learning is a combination of a traditional environment where learning takes place face-to-face between teachers and pupils and online learning, which allows pupils to use learning materials without physical presence (Fig. 2) [17].

The special advantages of blended learning are:

1. There is an opportunity to create a model of the educational process (for example, rotary, flexible, self-modified and in-depth virtual).

2. Asynchronous mode of operation.

3. The use of productive teaching methods - research methods, learning in small groups, business games, test technologies.

4. Organization of the system of control and self-control, initial and final control of knowledge.

5. Development and provision of meaningful learning in electronic form, creation of a basis for independent mastering of the course.

6. Combination of lecture classes with online training and networking (consultations, blogs, forums, chats).

There is a large number of blended learning models: simple, complex, more or less popular, and others. Most general secondary education institutions introduce more than one blended learning model for their pupils. In blended learning, both synchronous and asynchronous modes of communication can be used.

Synchronous mode involves interaction between the participants of distance learning, during which they are simultaneously in an electronic educational environment and communicate through audio and video conferencing.

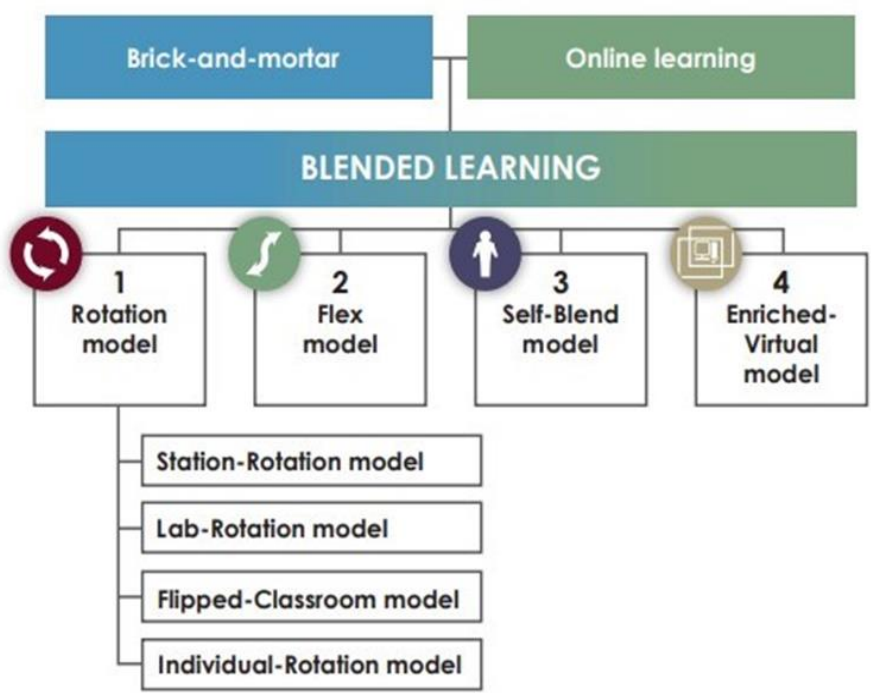

*Fig. 2. Taxonomy of blended learning [17]*

Asynchronous mode means the interaction between the participants of distance learning, in which they interact with each other with a delay in time, using interactive educational platforms, e-mail, forums, chats, social networks, and so on.

According to the Regulation on distance learning in general secondary education, approved by the order of the Ministry of Education and Science of Ukraine of September 8, 2020, No. 1115[18], in Ukraine at least 30 percent of the study time provided by the educational program is organized synchronously and asynchronously during quarantine restrictions. The teacher and the administration of the educational institution independently determine the model of a combination of synchronous and asynchronous modes of education, and it is possible to fill these components with educational activities in different ways. There are certain difficulties in the implementation of blended learning, which are considered from several sides in [10]:

Technical:
- providing participants with technologies - the step-by-step introduction of technologies from simple to more complex;
- resisting the desire to use technology simply because it is available is an important move.

Organizational:
- overcoming the prejudice that blended learning is not as effective as traditional classroom teaching;
- management and monitoring of progress – all elements of blended learning must be controlled and monitored.

Educational design:
- careful planning should determine at what stages cooperation is necessary, when to work online and when in the classroom;
- compliance with the best delivery environment to achieve the goal - it is necessary to carefully define the goals using Bloom's taxonomy and only then on this basis to determine the method of delivery;
- introduction of online interactive offers – after 10-20 minutes. Watching a video or participating in a webinar, the pupil must complete a learning activity;
- providing participants with requirements for performance evaluation in individual tasks and the course as a whole;
- ensuring coordination of all elements of the course – the guidelines should give answers to standard questions, contact information, schedule, and procedure for studying the course material.

However, there is currently a confusion of concepts. Due to the pandemic, most Ukrainian schools were forced to implement distance or blended learning. In blended learning there is a combination of distance and full-time work.

Therefore, virtual or augmented reality can be very relevant and necessary here. But there is a problem of correct justified use of these technologies. This requires techniques that should be developed and implemented in the learning process.

### 2.2. Practical use of virtual and augmented reality

According to study [1], virtual reality (VR) is an artificially created environment that can be accessed using special technology. A specific property of virtual reality is considered to be the maximum impact on most senses: touch, sight, smell, and hearing.

Augmented reality (AR) is artificial supplementing a real environment with extra information (images, audio, etc.). On the other hand, digital technology complements the real user environment, while virtual reality tries to completely transfer the user to an artificial environment.

I. Melnyk, N. Zaderei, and G. Nefodova in their study [1] in a separate section investigate mixed reality (MR), which combines augmented and virtual reality with the physical environment. In this reality, the boundaries between these concepts are blurred.

The active use of virtual VR and augmented AR reality in different areas forms a single reality (mixed reality, MR), in which the boundaries between augmented, virtual and physical worlds are blurred. Ericsson Consumer Lab's research came to this conclusion after examining the impact of AR and VR technologies on user habits and preferences.

The future of the physical real world will be shaped using the combined reality of MR, based on augmented (AR) and virtual reality (VR) [1].

Let us look at some applications of virtual and augmented reality that can be used in the process of blended learning.

MEL Chemistry VR is a course in chemistry lessons in virtual reality that corresponds to specific topics of the school curriculum, where virtual reality turns learning into an exciting process of learning the basics of chemistry, using scientific games and the method of immersion. Lessons are available in English only.

Anatomy VR is a mobile learning application that presents the anatomy to the user from a different perspective.

Using virtual reality, the user becomes part of one of the anatomical structures (observing the processes from the inside), having the opportunity to navigate along with these structures: blood, respiratory, digestive, urinary, lacrimal, and reproductive systems.

The application contains free access to some navigation routes in almost any of the mentioned systems. Additional content that can be purchased is offered in the section of the program sold through the application.

Anatomy VR can be used in two different modes: virtual reality and full-screen mode.

In virtual reality mode, a mobile device (smartphone) must be inserted into a virtual reality gadget to master certain learning material. The user can interact with navigation controls and anatomical information elements (by selecting one of them).

Although Anatomy VR provides the best virtual reality experience, the user can also enjoy and learn with this mobile application in full screen without the need for a virtual reality gadget.

Google Expeditions is an educational tool that allows you to travel in the virtual world and explore objects in augmented reality. There are modes of studying historical monuments, exploring objects at the level of atoms, swimming near sharks, or going into outer space.

In Google Expeditions, a teacher becomes a guide who goes on a video tour with a class or group or shows augmented reality objects. They can use special tools for detailed study of individual subjects. You do not have to be part of a group to immerse yourself in virtual or augmented reality, and users can do it on their own. The application contains the following functions:

− Explore the world around you with your phone, tablet, or virtual reality device.
− The teacher will be able to become a guide for small groups of users who connect to your expedition from their mobile devices.
− You can join other guides' expeditions to explore the virtual world and explore objects in augmented reality. To do this, use, for example, glasses Google Cardboard or Daydream.
− Connect your device to the same Wi-Fi network as the group. If the guide downloaded the expedition, you do not need an Internet connection to participate.
− Choose expeditions from an extensive catalog - more than 800 virtual or augmented reality video tours. Each of them consists of carefully selected materials and is provided with descriptions, information blocks, and questions.

EON-XR is an augmented or virtual reality application equipped with features for distance learning and training in a practical environment.

Lessons and training can be conducted in AR or VR modes either individually or in groups, providing all the requirements for teachers, pupils, or groups of pupils.

With the EON-XR, users can quickly create compelling content on their phones, tablets, computers, and headsets using the EON Reality library with more than 1 million digital assets, as well as importing or buying assets from other sources. The main characteristics of the service are:
- creation, publication, demonstration, recording, and evaluation of lessons and their content using a smartphone or tablet;
- gathering teachers and pupils for virtual interaction with the lesson in a safe virtual space;
- intuitive platform without code, designed to encourage pupils and teachers to create content for deeper learning and collaboration;
- embedded asynchronous video with 3D models and 360° environments to provide self-learning anytime, anywhere;
- evaluating the effectiveness of pupils or groups of pupils through assessment and quizzes to effectively track the level of mastery of educational material.

The following tools are available:
- Availability of several regions for the EON-XR meeting.
- Playlist function.
- Pupil assessment function.
- Create Express 360 lessons.
- General optimization, including playback mode.
- There is a Russian localization.

**2.3. Technological component of the use of virtual reality services**

Types of virtual reality are identified in the publication [19]. The authors identify the following types of virtual reality:
- VR technologies with the effect of full immersion, providing a plausible simulation of the virtual world with a high degree of detail. Their implementation requires a high-performance computer capable of recognizing user actions and responding to them in real-time, and special equipment that provides the effect of immersion;
- VR technologies without immersion. These include simulations with images, sound, and controllers that are broadcast on the screen, preferably widescreen. Such systems are considered virtual reality, because, in terms of the degree of impact on the viewer, they by far surpass other means of multimedia, although they do not fully implement the requirements of VR;
- VR technologies with common infrastructure. These include Second Life - a three-dimensional virtual world with elements of a social network, which has more than a million active users, the game Minecraft and others. Such worlds do not provide full immersion (however, Minecraft already has a virtual reality version that supports the Oculus Rift and Gear VR helmets). But in virtual worlds, interaction with other users is well organized, which is often lacking in the products of "true" virtual reality;
- Virtual worlds are used not only in the gaming industry: thanks to platforms such as 3D Immersive Collaboration, you can organize work and study 3D spaces - this is called "collaboration with the effect of presence." Ensuring full immersion and, at the same time, user interaction in virtuality is one of the important areas of VR development;
- VR based on Internet technologies. These include primarily Virtual Reality Markup Language, similar to HTML. Now, this technology is considered obsolete, but it is possible that in the future virtual reality will be created, including the use of Internet technology.

It is known that a person receives 80% of the information through sight. Therefore, the developers of BP systems pay great attention to the devices that provide image formation. As a rule, they are supplemented by stereo devices, work on tactile influences, and even imitation of odors. The effect on taste buds has not yet been reported [19].

Virtual reality helmet. Modern virtual reality helmets (HMD-display, head-mounted display, video helmet) contain one or more displays that display images for the left and right eyes, a system of lenses to adjust the geometry of the image, as well as a tracking system that tracks the orientation of the device in space. In appearance, they now look like glasses, so they are increasingly called VR headsets (VR headsets) or just virtual reality glasses. They can be divided into three groups:

1. Glasses in which image processing and output are provided by a smartphone (Android, iPhone, Windows Phone). Modern smartphone is a high-performance device capable of processing three-dimensional images. Smartphone displays have a fairly high resolution. Almost every smartphone is equipped with sensors that allow you to determine the position of the device in space.

2. Glasses in which image processing is provided by an external device (PC, Xbox, PlayStation, etc.). The external device must be high-performance, and the glasses are equipped with position sensors.

3. Autonomous virtual reality glasses (Lenovo Mirage Solo, together with Google, Oculus Quest from Facebook, Samsung Gear VR, etc.).

Helmets are the main component of VR with full immersion, as they not only provide a three-dimensional image and stereo sound but also partially isolate the user from the surrounding reality [19].

MotionParallax3D displays. Such displays use the inherent mechanism of human perception of volume-motion parallax. To do this, at each point in time for the viewer, based on its position relative to the screen, a corresponding projection of the three-dimensional object is generated. Moving around the scene, the user can view it from all sides, while all objects of the scene will move relative to each other. The phenomenon of parallax repeatedly enhances the perception of volume. Unlike 3D cinema and 3D TV, which use only binocular vision, MotionParallax3D technology allows the user to view the 3D scene from all sides as if all its objects were real. Shifting the viewer relative to the screen, which violates the effect of volume in 3D cinema, in the MotionParallax3D system only enhances the effect. A system that uses a parallax mechanism must capture the smallest movements of the user's head and track them with high speed and accuracy so that the brain does not detect distortions in the geometry of objects caused by delayed image changes.

Virtual reality gloves (information gloves, datagloves). These gloves have sensors that allow you to monitor the movement of the wrists and fingers. Technically, this can be done in a variety of ways: using fiber optic cables, strain gauges or piezoelectric sensors, and electromechanical devices (such as potentiometers). For example, scientists at EPFL and ETH Zurich have developed ultralight gloves (weighing less than 8 grams each finger and only 2 mm thick). They provide extremely realistic tactile feedback and thus give unprecedented freedom of movement [19].

Virtual reality costume. This suit should monitor the change of position of the whole body of the user and transmit tactile, temperature, and vibration sensations, and in combination with a helmet - visual and auditory.

Control devices. To interact with the virtual environment, special joysticks (gamepads, wands) are used, which contain built-in position and motion sensors, as well as buttons and scroll wheels, like a mouse. Now such joysticks are increasingly made wireless [19].

In Ukraine, the above mentioned technologies are used in small quantities today, because these devices and their software are too expensive for most educational institutions. Therefore, the use of virtual and auxiliary reality technologies in the educational process is poorly understood, but nevertheless relevant for further research.

## 2.4. Recommendations for the use of augmented and virtual reality in the process of blended learning

Blended learning is a solution for education in an era of revolutions and an ideal learning model for experiments with AI, VR, and AR. Teachers and pupils are adapting to the new forms of distance education and the difficulties that arise with it, but it is also a great opportunity to explore interactive technologies that are ideal for virtual learning. Modern teachers explore and experiment with ways to incorporate these technologies into practices that enrich the educational experience for all participants.

Augmented reality is an integral part of the educational process of the future, characterized by a combination of virtual technology and the real world. AR technologies make it possible to explain abstract concepts, certain theories, or things that cannot be represented. One of the distinctive features is the conversion of 2D images into 3D to make them look realistic and improve the existing environment with animation and sounds. The effectiveness of this format of learning has been proven: pupils perceive and remember visual images much faster. The use of such interactive technologies in class increases the motivation to learn.

It is noted [20] that among the most promising elements of this technology is its ease of use and continuous integration into the curriculum. Teachers can simply insert QR codes into their learning materials (such as PowerPoint slides, LMS, or handouts) to give pupils instant access. Pupils can then scan the code using their phones or tablets to liven up the AR experience in their own homes or view an interactive 3D model from their desktop, laptop, or Chromebook. You do not need expensive hardware like headsets, or complicated software or applications.

With the help of augmented reality, you can expand the opportunities for learning in any subject of the school curriculum. The teacher can also suggest that the pupils continue to study the material independently, doing interactive homework.

Let's outline the possibilities of using augmented or virtual reality during blended learning:

1. Visualization of educational material and diversification of the educational process. Abstract topics and concepts can be made more interesting and understandable. After all, most mistakes pupils make when they do not fully understand all the properties of what they are currently studying. A flat image cannot be held or viewed from all sides. Augmented and virtual reality can be a great opportunity to master complex topics.

2. Organization of group or project work. In this case, it will not be a formalized division into groups or micro-groups of pupils in the class, but full-fledged teamwork. In this case, the result of certain tasks depends on each participant. Project work involves solving a problem by a pupil or a group of pupils, which includes, on the one hand, the use of various methods, teaching aids, and on the other - the integration of knowledge and skills from different fields of science, technology, creativity. Pupils have the opportunity not only to use knowledge of the discipline but also to learn to negotiate and make decisions together, to be responsible according to their role in the educational team and to interpret the results of their activities. And the teacher acts as a tutor, mentor, and team leader (leader) and is a full member of the group of pupils.

3. Use of the most modern technologies. Augmented and virtual reality technologies should be used in cases where it is most difficult to understand the training material. At the same time, it is necessary to consider all possibilities of their use in advance. Learning material should be available to every pupil, free and understandable.

4. Additional tools for assessing academic achievement. Independent work, control, or test task can be organized using augmented or virtual reality (for example, in the form of a quest or performing tasks in the form of a game). This format is interesting and encourages the knowledge of additional facts, a deeper understanding of the subject.

As these and other forms of new technologies are introduced into education, blended learning options will continue to expand, leading to the formation of the class of the future, which provides an open educational platform with interdisciplinary and innovative approaches using digital technologies that improve the educational process focused on pupils. And this is convincing because the emerging technologies have the potential to enrich education. Therefore, a model was designed to use augmented and virtual reality in blended learning in general secondary education (Fig. 3).

The main purpose of using virtual and augmented reality in blended learning is to acquire knowledge and practical skills, form digital and research competence of students. The teacher can choose a convenient model of blended learning, prepare the necessary materials and teaching aids to effectively combine online learning with classroom learning.

When organizing blended learning in a group or individually, the teacher uses in addition to standard technical means also means of AR and VR. The method of training is chosen depending on the lesson plan: lecture-monologue, modeling, brainstorming.

During the study, the student acquires knowledge, skills and abilities in a particular subject. The result of blended learning involving AR and VR is the acquisition of digital and research competencies in different formats of learning.

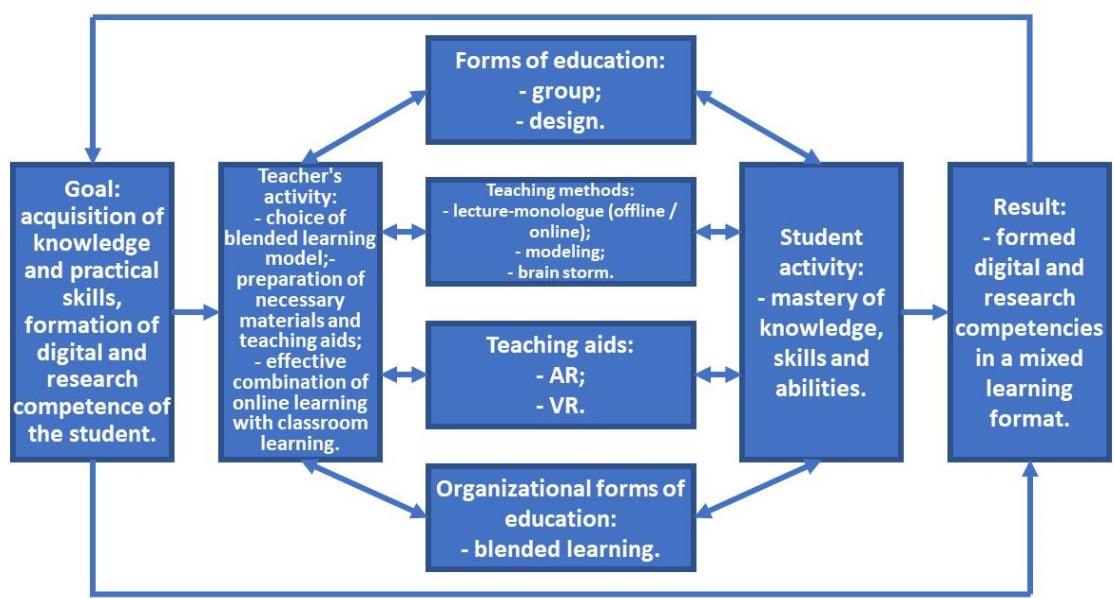

*Fig. 3. The model to use augmented and virtual reality in blended learning in general secondary education*

Based on the designed model, the methodology of using augmented and virtual reality in blended learning in general secondary education was developed.

The structure of the methodology of using augmented and virtual reality in blended learning in general secondary education:

*Target component.*

Purpose: acquisition of knowledge and practical skills, formation of digital and research competence of the pupils.

Target group: teachers, pupils.

*Content component.*

Teacher's activity: choice of model of blended learning; preparation of necessary materials and teaching aids; an effective combination of online learning with classroom learning.

Pupils' activity: mastering knowledge, skills and abilities.

*Technological component.*

Teaching methods: lecture-monologue (offline/online); modeling; brain storm.

Forms of training: group, project.

Learning tools: augmented and virtual reality tools.

Organizational forms of learning: blended learning.

*Resultant component:* formed digital and research competencies in a mixed learning format.

As can be seen, the methodology is quite universal and can be used for any subject in general secondary education. The key mode of learning is blended learning. This is due to the use of augmented and virtual reality tools. The use of these teaching aids must be combined with traditional ones. For example, theoretical material may be effectively presented using augmented and virtual reality. The flexibility of the methodology is provided by the variability of the use of one or another tool of augmented and virtual reality.

Here are the examples of augmented or virtual reality software products:
− for physics and chemistry: MEL Chemistry VR;
− for anatomy: Anatomy VR;
− for geography: Google Expeditions;
− for the organization of the educational process using AR, VR: EON-XR.

There are many possible ways of using augmented or virtual reality in general secondary education, but the purpose of our study is not to describe all the available tools.

It is better to plan the lesson using separate tools from AR and VR, because full immersion in virtual reality will lead to a change in the organizational form of learning (distance learning).

The use of separate AR and VR tools is best used for the following types of lessons:
− a lesson of learning new knowledge;
− lesson of formation of abilities and skills;
− lesson of generalization and systematization;
− combined lesson.

It is best to use AR and VR tools to support the following processes:
− pupils' perception and awareness of factual material (lesson of learning new knowledge, combined lesson);
− understanding the connections and dependencies between the elements of the learning (lesson of learning new knowledge, combined lesson);
− creative transfer of knowledge and skills to new situations (lesson of formation of abilities and skills);
− generalization and systematization of basic theoretical principles and relevant ideas of science (lesson of generalization and systematization).

The use of AR and VR is considered not only in theoretical terms, but also has a practical implementation. The teachers' readiness to use augmented reality in the educational process was studied in [21]. The use of gamification and modeling as cognitive technologies, as well as social networks as a synthetic environment for social development is described in [22].

In 2021, the Institute of Information Technologies and Learning Tools of the National Academy of Pedagogical Sciences of Ukraine launched a study "The use of digital technologies in the process of blended learning in general secondary education" and obtained the following results:

- developed criteria and indicators for the selection of digital technologies in the process of blended learning;
- substantiated and developed methods of using digital technologies in the process of blended learning in general secondary education;
- developed a method of using augmented and virtual reality in blended learning in general secondary education.

The results of the research were implemented and approved within the following scientific and practical events:

January 27, 2021 – Webinar on the topic: "Information technology in science and education", Vinnytsia Mykhailo Kotsyubynsky State Pedagogical University, Vinnytsia.

February 11, 2021 – Reporting scientific conference of the Institute of Information Technologies and Teaching Aids of the National Academy of Pedagogical Sciences of Ukraine, Kyiv.

February 24, 2021 – Master class "Preparation of mathematics teachers for blended and distance learning of students", Kryvyi Rih State Pedagogical University, Kryvyi Rih.

April 7, 2021 – All-Ukrainian scientific seminar "Security of the individual in the digital environment: legal, psychological and technological aspects", State Research Institute of the Ministry of Internal Affairs of Ukraine, Kyiv.

August 19, 2021 – All-Ukrainian August Forum "Education of Ukraine 30 without barriers: vectors of quality and success", strategic session "NEW UKRAINIAN SCHOOL: innovative dimensions of teaching, learning, and education".

The empirical testing showed an increase in the ICT competence of learners during the proposed training and master classes.

## 5. CONCLUSIONS AND PROSPECTS FOR FURTHER RESEARCH

Thus, in the context of the digital transformation of education in Ukraine, the use of digital technologies in the educational process opens wide prospects for improving the efficiency of the educational process. And the orientation of educational institutions towards modern pedagogical trends will help to increase the motivation of learners, in particular, the development of their digital and research competence.

We believe that virtual augmented and mixed reality technologies have a wide range of impact on human perception of the surrounding, in particular, the use of the above mentioned technologies will significantly enrich the educational process and contribute to the formation of pupils' digital and research competence.

The use of augmented and virtual reality technologies provides pupils with new opportunities and prospects focused on the acquisition of practical skills, promotes the development and self-education of each pupil, gives them the opportunity to gain the latest knowledge and get practical training. The use of AR and VR technologies brings science closer to life, recreates real-life situations, helps to create artificial spaces for unsolved problems. This creates new opportunities for mastering practical skills, provides research experience, makes learning a vivid process, prevents distractions from learning and increases motivation to learn, helps to better understand the set of concepts, definitions, theorems, properties that pupils deal with when learning certain topics.

The study described the special benefits of blended learning. It has been found that there are different models of blended learning. The model of using augmented and virtual reality in blended learning in general secondary education is proposed. It consists of the following blocks: goal; teacher's activity; forms of education; teaching methods; teaching aids; organizational forms of education; pupil activity and result. Based on the model, the methodology of using augmented and virtual reality in blended learning in general secondary education was developed. The methodology is quite universal and can be used for any subject in general secondary education. The key organizational form of learning is blended learning. The flexibility of the technique is provided by the variability of the use of one or another tool of augmented and virtual reality. It is better to plan the lesson using separate tools from AR and VR, because full immersion in virtual reality will lead to a change in the organizational form of learning (distance learning).

The possible ways to use augmented reality suggest:
− the need to transform expensive technologies into budget options suitable for smartphones and simple computers;
− augmented reality technologies should be aimed at acquiring skills, transferring and controlling knowledge.

We see the prospects of further research in the elaboration of guidelines on the use of augmented reality in the process of blended learning in general secondary education.

**FINANCING**

The article presents the results of the study "Use of digital technologies in the process of blended learning in general secondary education", the winner of the competitive selection for 2021 applied research to support young scientists working (studying) in subordinate institutions of NAES of Ukraine (Resolution of the Presidium of NAES of Ukraine of February 1, 2021, No 1-2 / 2-13). The authors of the article are performers of the applied research.

# ВИКОРИСТАННЯ ІНСТРУМЕНТІВ ДОПОВНЕНОЇ ТА ВІРТУАЛЬНОЇ РЕАЛЬНОСТІ В ЗАКЛАДІ ЗАГАЛЬНОЇ СЕРЕДНЬОЇ ОСВІТИ В УМОВАХ ЗМІШАНОГО НАВЧАННЯ

**Коваленко Валентина Володимирівна**
кандидат педагогічних наук,
старша наукова співробітниця відділу хмаро орієнтованих систем інформатизації освіти
Інститут інформаційних технологій і засобів навчання НАПН України, м. Київ, Україна
ORCID ID 0000-0002-4681-5606
*valyavako@gmail.com*

**Мар'єнко Майя Володимирівна**
кандидат педагогічних наук,
старша наукова співробітниця відділу хмаро орієнтованих систем інформатизації освіти
Інститут інформаційних технологій і засобів навчання НАПН України, м. Київ, Україна
ORCID ID 0000-0002-8087-962X
*popelmayal@gmail.com*

**Сухіх Аліса Сергіївна**
кандидат педагогічних наук,
старша наукова співробітниця відділу хмаро орієнтованих систем інформатизації освіти
Інститут інформаційних технологій і засобів навчання НАПН України, м. Київ, Україна
ORCID ID 0000-0001-8186-1715
*alisam@ukr.net*

**Анотація.** У статті досліджується проблема використання доповненої та віртуальної реальності під час змішаного навчання в закладі загальної середньої освіти. Аналіз останніх досліджень та публікацій показав, що використання доповненої та віртуальної реальності в навчальному процесі вже розглядалося вченими. Однак цільовою групою у цих дослідженнях є студенти закладів вищої освіти. Водночас більшість праць науковців присвячена проблемі впровадження доповненої реальності у традиційний освітній процес. Однак використання технології доповненої та віртуальної реальності під час змішаного навчання залишається практично не дослідженим. У проведеному дослідженні аналізується зміст поняття "змішане навчання". Розглянуто концептуальні принципи змішаного навчання. Було виявлено, що в поняття "змішане навчання" вчені привносять різні трактування, тому їх визначення дещо відрізняються. Іноді дослідники розрізняють компоненти змішаного навчання: очне та онлайн навчання. Дослідження розкриває особливості та переваги змішаного навчання та принципи організації змішаного навчання. Було виявлено, що існують певні труднощі у впровадженні змішаного навчання. У статті викладено практичне використання віртуальної та доповненої реальності. Дано визначення доповненої та віртуальної реальності. Змішана реальність розглядається як окремий вид поняття «віртуальна реальність». Розглядається використання окремих додатків віртуальної та доповненої реальності, які можна застосовувати під час змішаного навчання (MEL Chemistry VR; Anatomyou VR; Google Expeditions; EON-XR). У результаті дослідження автори наводять можливі способи використання доповненої реальності в навчальному процесі. Була спроєктована модель використання доповненої та віртуальної реальності у змішаному навчанні в загальній середній освіті. Вона складається з таких блоків: мета; діяльність учителя; форми навчання; методи навчання; засоби навчання; організаційні форми навчання; активність учнів та результат. На основі моделі розроблено методику використання доповненої та віртуальної реальності у змішаному навчанні в закладі загальної середньої освіти. Методика містить такі компоненти: цільовий, змістовий, технологічний, результативний. Методика універсальна і може бути використана для будь-якого предмета загальної середньої освіти. Визначено типи уроків, на яких доцільно використовувати доповнену (AR) та віртуальну реальність (VR). Дано рекомендації, на якому етапі уроку краще використовувати інструменти AR та VR (залежно від типу уроку).

**Ключові слова:** змішане навчання; доповнена реальність; віртуальна реальність; заклади загальної середньої освіти.

# ИСПОЛЬЗОВАНИЕ ИНСТРУМЕНТОВ ДОПОЛНЕННОЙ И ВИРТУАЛЬНОЙ РЕАЛЬНОСТИ В СРЕДНЕМ ОБЩЕОБРАЗОВАТЕЛЬНОМ УЧЕБНОМ ЗАВЕДЕНИИ В УСЛОВИЯХ СМЕШАННОГО ОБУЧЕНИЯ

**Коваленко Валентина Владимировна**
кандидатпедагогических наук,
старший научный сотрудник отдела облачно ориентированных систем информатизации образования
Институт информационных технологий и средств обучения НАПН Украины, г. Киев, Украина
ORCID ID 0000-0002-4681-5606
*valyavako@gmail.com*

**Марьенко Майя Владимировна**
кандидат педагогических наук,
старший научный сотрудник отдела облачно ориентированных систем информатизации образования
Институт информационных технологий и средств обучения НАПН Украины, г. Киев, Украина
ORCID ID 0000-0002-8087-962X
*popelmayal@gmail.com*

**Сухих Алиса Сергеевна**
кандидат педагогических наук,
старший научный сотрудник отдела облачно ориентированных систем информатизации образования
Институт информационных технологий и средств обучения НАПН Украины, г. Киев, Украина
ORCID ID 0000-0001-8186-1715
*alisam@ukr.net*

**Аннотация.** В исследовании рассматривается проблема использования дополненной и виртуальной реальности в процессе смешанного обучения в средних общеобразовательных учреждениях. Анализ последних исследований и публикаций показал, что использование дополненной и виртуальной реальности в образовательном процессе уже рассматривалось учеными. Однако целевая группа в этих исследованиях – студенты высших учебных заведений. При этом большая часть работ ученых посвящена проблеме внедрения дополненной реальности в традиционный образовательный процесс. Однако, использование технологий дополненной и виртуальной реальности в процессе смешанного обучения остается практически неизученным. В исследовании анализируется значение понятия «смешанное обучение». Рассмотрены концептуальные принципы смешанного обучения. Показано, что ученые рассматривают разные трактовки к пониманию термина «смешанное обучение», поэтому их определения несколько различаются. Иногда исследователи различают компоненты смешанного обучения: очное и онлайн-обучение. В исследовании представлены преимущества смешанного обучения и модели смешанного обучения. Исследование показало, что существуют некоторые трудности при внедрении смешанного обучения. В статье речь идет о практическом использовании виртуальной и дополненной реальности. Дано определение дополненной и виртуальной реальности. Смешанная реальность рассматривается как отдельное понятие. Рассмотрены некоторые приложения виртуальной и дополненной реальности, которые можно использовать в процессе смешанного обучения (MEL Chemistry VR; Anatomyou VR; Google Expeditions; EON-XR). В результате исследования авторы приводят возможные способы использования дополненной реальности в учебном процессе. Была спроектирована модель использования дополненной и виртуальной реальности в смешанном обучении в среднем общеобразовательном учреждении. Она состоит из следующих блоков: цель; педагогическая деятельность; формы обучения; методы обучения; средства обучения; организационные формы обучения; ученическая активность и результат. На основе модели разработана методика использования дополненной и виртуальной реальности в смешанном обучении в среднем общеобразовательном учреждении. Методика содержит следующие компоненты: целевой, содержательный, технологический, результативный. Методика достаточно универсальна и может применяться к любому предмету общего среднего образования. Определены виды уроков, в которых целесообразно использовать дополненную (AR) и виртуальную (VR) реальность. Даны рекомендации, на каком этапе урока лучше использовать инструменты AR и VR (в зависимости от типа урока).

**Ключевые слова:** смешанное обучение; дополненная реальность; виртуальная реальность; средние общеобразовательные учреждения.